%% file: main.tex
\documentclass[sigconf,anonymous=False]{acmart}

\usepackage{graphicx}
\usepackage{caption}
\usepackage{subcaption}
\usepackage{xcolor}
\usepackage{algorithm}
\usepackage{algpseudocode}
\usepackage{xcolor,colortbl}

\definecolor{Gray}{gray}{0.85}
\definecolor{LightCyan}{rgb}{0.88,1,1}
\newcolumntype{a}{>{\columncolor{Gray}}p}

\AtBeginDocument{%
  \providecommand\BibTeX{{%
    \normalfont B\kern-0.5em{\scshape i\kern-0.25em b}\kern-0.8em\TeX}}}

\setcopyright{acmlicensed}
\copyrightyear{2024}
\acmYear{2024}
\acmDOI{XXXXXXX.XXXXXXX}

\acmConference[Arxiv.org]{2024}{}

\acmISBN{978-1-4503-XXXX-X/18/06}

\begin{document}

\title{Acoustic Side Channel Attack on Keyboards Based on Typing Patterns}


\author{Alireza Taheritajar}
\email{ataheritajar@augusta.edu}
\orcid{0009-0002-5642-6329}
\affiliation{%
  \institution{Augusta University}
  \streetaddress{1120 15th St.}
  \city{Augusta}
  \state{GA}
  \country{USA}
  \postcode{30912}
}

\author{Reza Rahaeimehr}
\email{rrahaeimehr@augusta.edu}
\orcid{0000-0003-0305-3661}
\affiliation{%
  \institution{Augusta University}
  \streetaddress{1120 15th St. }
  \city{Augusta}
  \state{GA}
  \country{USA}}

\renewcommand{\shortauthors}{A. Taheritajar, et al.}

\begin{abstract}
 Acoustic side-channel attacks on keyboards can bypass security measures in many systems that use keyboards as one of the input devices. These attacks aim to reveal users' sensitive information by targeting the sounds made by their keyboards as they type. Most existing approaches in this field ignore the negative impacts of typing patterns and environmental noise in their results. This paper seeks to address these shortcomings by proposing an applicable method that takes into account the user's typing pattern in a realistic environment. Our method achieved an average success rate of 43\% across all our case studies when considering real-world scenarios.
\end{abstract}





\settopmatter{printfolios=true}
\maketitle

\input{introduction}

\input{related_work}

\input{our_approach}

\input{Experimental_analysis}

\input{Conclusion}

\bibliographystyle{ACM-Reference-Format}
\bibliography{references}


\end{document}

%% file: introduction.tex
\section{Introduction}
\label{introduction}

\begin{figure}[htbp]
    \includegraphics[width=0.45\textwidth]{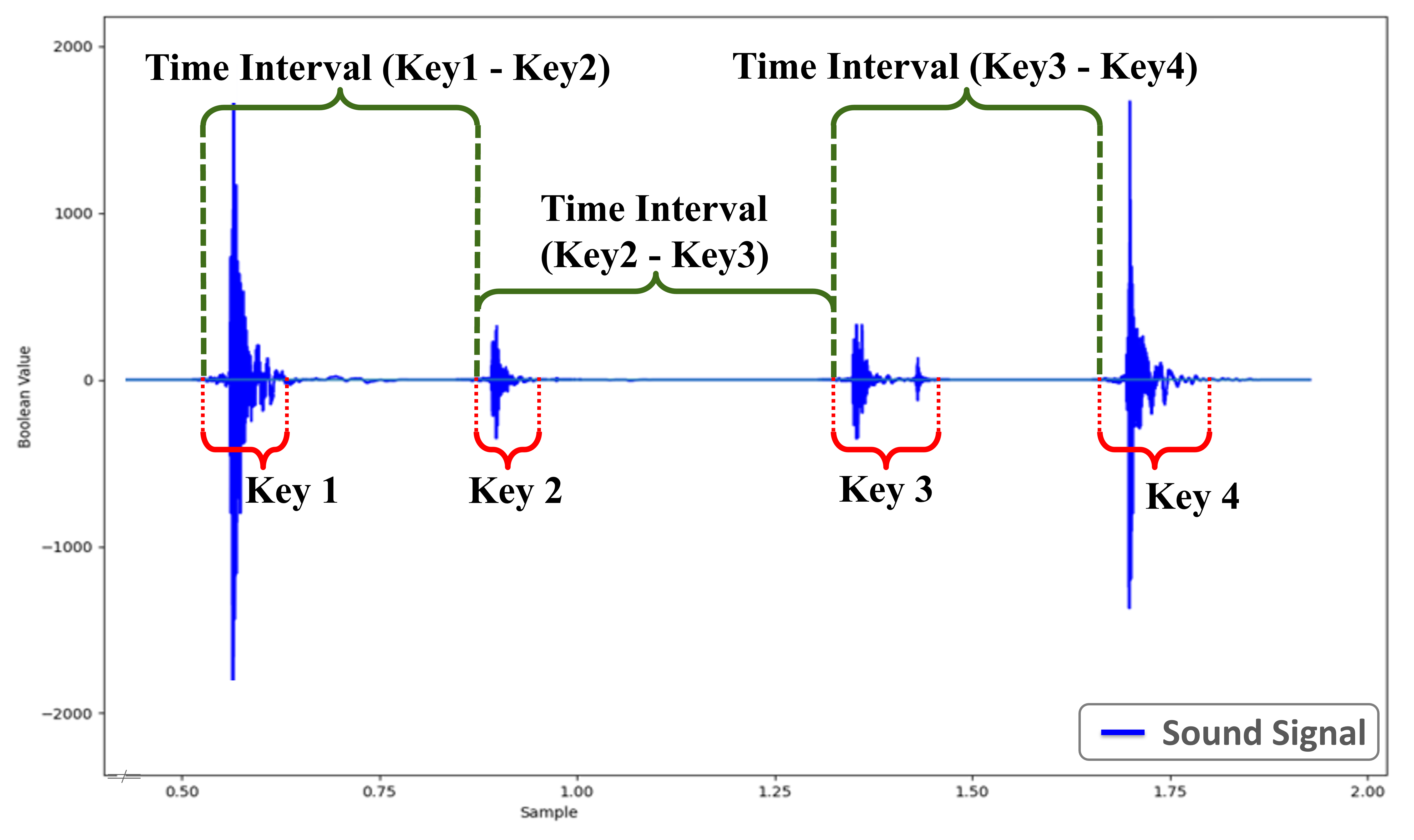}
    \centering
    \caption{Sample of acoustic waves produced by typing a four-letter word}
    \label{fig:Method}
\end{figure}

There has been a growing concern over personal data and information security in recent years. With the increasing use of digital devices and online communication, protecting sensitive data has become crucial to modern life. The process of acquiring knowledge about a system by analyzing its side effects is commonly referred to as a Side-Channel Attack~\cite{taheritajar2023survey}. Electronic device emissions have long been a concern within the security and privacy communities~\cite{briol1991emanation}, as they are susceptible to side-channel attacks. Acoustic side-channel attacks \cite{asonov2004keyboard,backes2010acoustic} can pose a significant threat to digital security. This type of attack takes advantage of unintended sound emissions generated by electronic devices to extract sensitive information. One of the threats to digital security involves keyboard acoustic side-channel attacks \cite{berger2006dictionary}, where malicious actors can extract sensitive information, such as emails, passwords, or credit card details \cite{ponnam2013keyboard}. Such attacks can be hazardous, as they can bypass traditional security measures like passwords \cite{ross2005stronger,yan2004password} and encryption. These attacks usually analyze the sound of each individual key in a time or frequency domain\cite{zhuang2009keyboard}.

One of the most concerning aspects of keyboard acoustic side-channel attacks is that they can be conducted remotely without the need for physical access to the target device \cite{compagno2017don, taheritajar2023survey}. This makes them an attractive target for cyber-criminals, who can easily reach numerous users. To avoid this type of attack, it is essential to guarantee that sound cannot be captured in the environment due to the high efficiency of modern microphones \cite{zhuang2009keyboard}. Attackers can analyze the sound waveforms of keystrokes to determine keystrokes' timing, intensity, and frequency, exposing sensitive information. One of the major difficulties in carrying out keyboard acoustic side-channel attacks is accurately capturing the sound of keystrokes. This entails using high-quality microphones and advanced signal-processing techniques to eliminate background noise and extract the sound of keystrokes. Once the sound has been captured, attackers can use various data analysis techniques, including statistical analysis, machine learning algorithms, signal processing techniques \cite{taheritajar2023survey}, acoustic triangulation attack\cite{fiona2006keyboard}, and Time Difference of Arrival (TDoA) \cite{zhu2014context}. 

Some past studies have demonstrated significant success with their methodologies. However, most of them had to restrict the environmental conditions or neglect the distinct features in their experiments to avoid any interference that could impact the outcomes. For instance, they usually ignored the effect of environmental noise or users' typing patterns on their research \cite{taheritajar2023survey}. Recognizing typing patterns is a complex task due to variations in how people use keys\cite{halevi2012closer}.
This variability adds to the challenge of achieving reliable and accurate recognition, as highlighted in Asonov and Agrawal's study\cite{asonov2004keyboard}. Furthermore, it is well-established that most attributes of emanations exhibit non-uniform patterns across different device models and are often affected by environmental factors. Another important factor that can disrupt existing methods is the keyboard model used by the user. Usually, when the keyboard model is changed, most algorithms can no longer maintain their previous predictive power due to the unique sound characteristics produced by each keyboard. In addition, recently proposed approaches, especially those that use deep learning networks, have added complexity to obtaining stable results during training, performance, or execution. So, it is crucial to propose a novel method that covers all of these disadvantages in this field of research.
This paper explores the concept of keyboard acoustic side-channel attacks and their potential to compromise digital security and proposes a method that covers aforementioned problems and weaknesses. We keep our proposed method simple and applicable to realistic scenarios. The process involves capturing and analyzing audio recordings of keystrokes, extracting inter-keystroke time intervals, and keypress timings (Figure \ref{fig:Method}). Then, we train a  statistical model using this data to predict the user's keystrokes. The model is then tested on unknown recorded sounds of the keystrokes of the same user to determine its accuracy and effectiveness. Finally, we enhance our results by filtering the wrongly predicted words by an English dictionary. By using this approach, we can analyze the typing patterns of users and use this information to predict the texts they type, even in real environments with ambient noise. We don't restrict our users to specific keyboard models.

The experiment results indicate that a user's typing pattern in keyboard acoustic side-channel attacks can be used to bypass security countermeasures. We collected confidential data from 20 users' typing samples to study the effect of letter count and typing patterns on success rate. We were able to achieve a success rate of approximately 43\% on average across all of our test cases. The threat of keyboard acoustic side-channel attacks emphasizes the significance of having robust cybersecurity practices. By being well-informed about the risks of acoustic side channel attacks and implementing appropriate measures to protect sensitive data, individuals and organizations can enhance their security against this threat.

The article is structured as follows: First, we provide a brief overview of previous research on keyboard emanation attacks in section \ref{Related Work}. Then, we present our newly proposed method in section \ref{Implementation}. In section \ref{Experimental Analysis}, we provide experimental results, while section \ref{Limitations} covers the limitations and countermeasures. Finally, we conclude the paper and propose potential future research in section \ref{Conclusion}.

%% file: related_work.tex
\begin{figure*}
     \centering
     \begin{subfigure}[b]{0.33\textwidth}
         \centering
         \includegraphics[width=\textwidth]{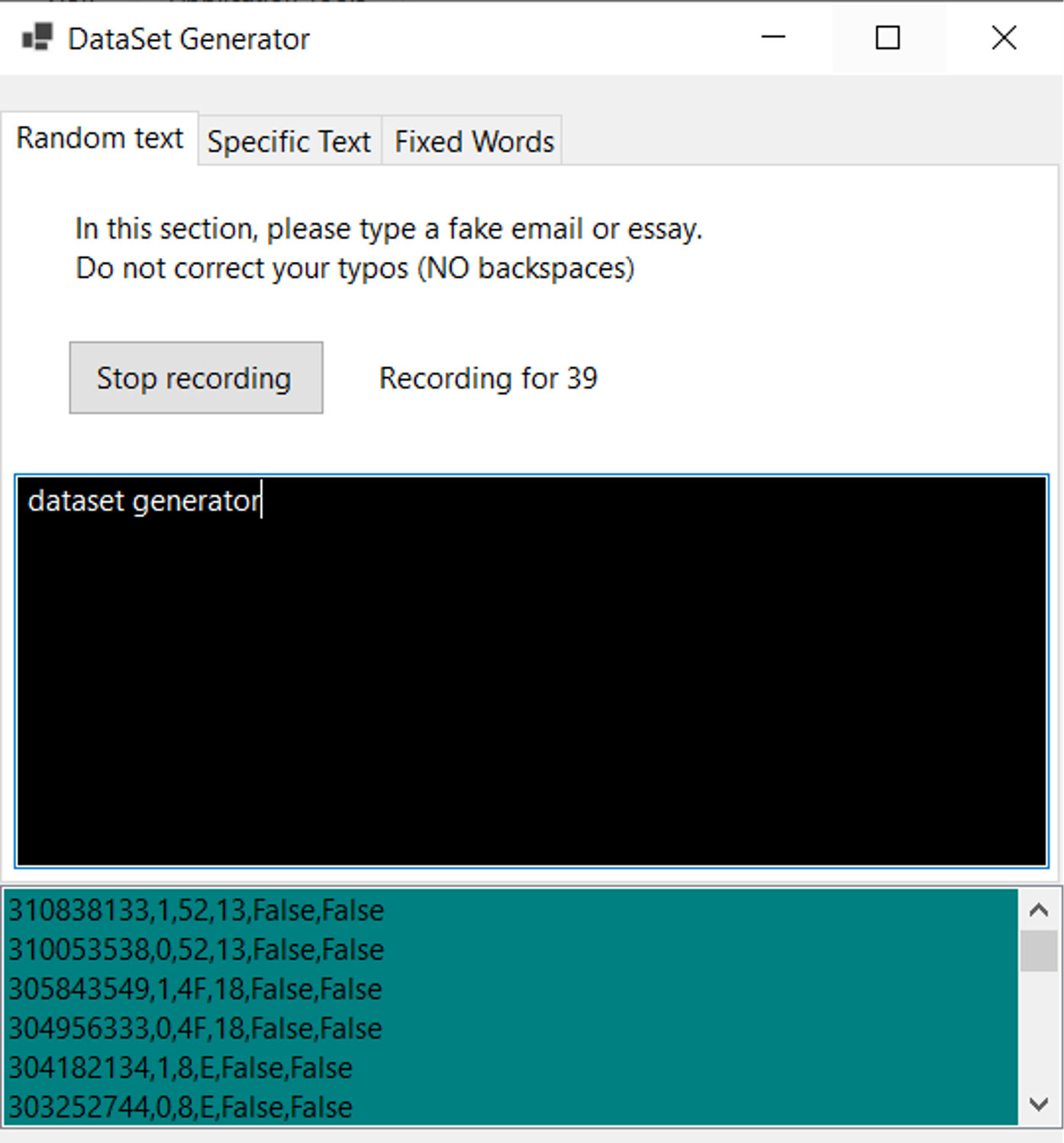}
         \caption{Random Text} \label{fig:random}
     \end{subfigure}
     \hfill
     \begin{subfigure}[b]{0.33\textwidth}
         \centering
         \includegraphics[width=\textwidth]{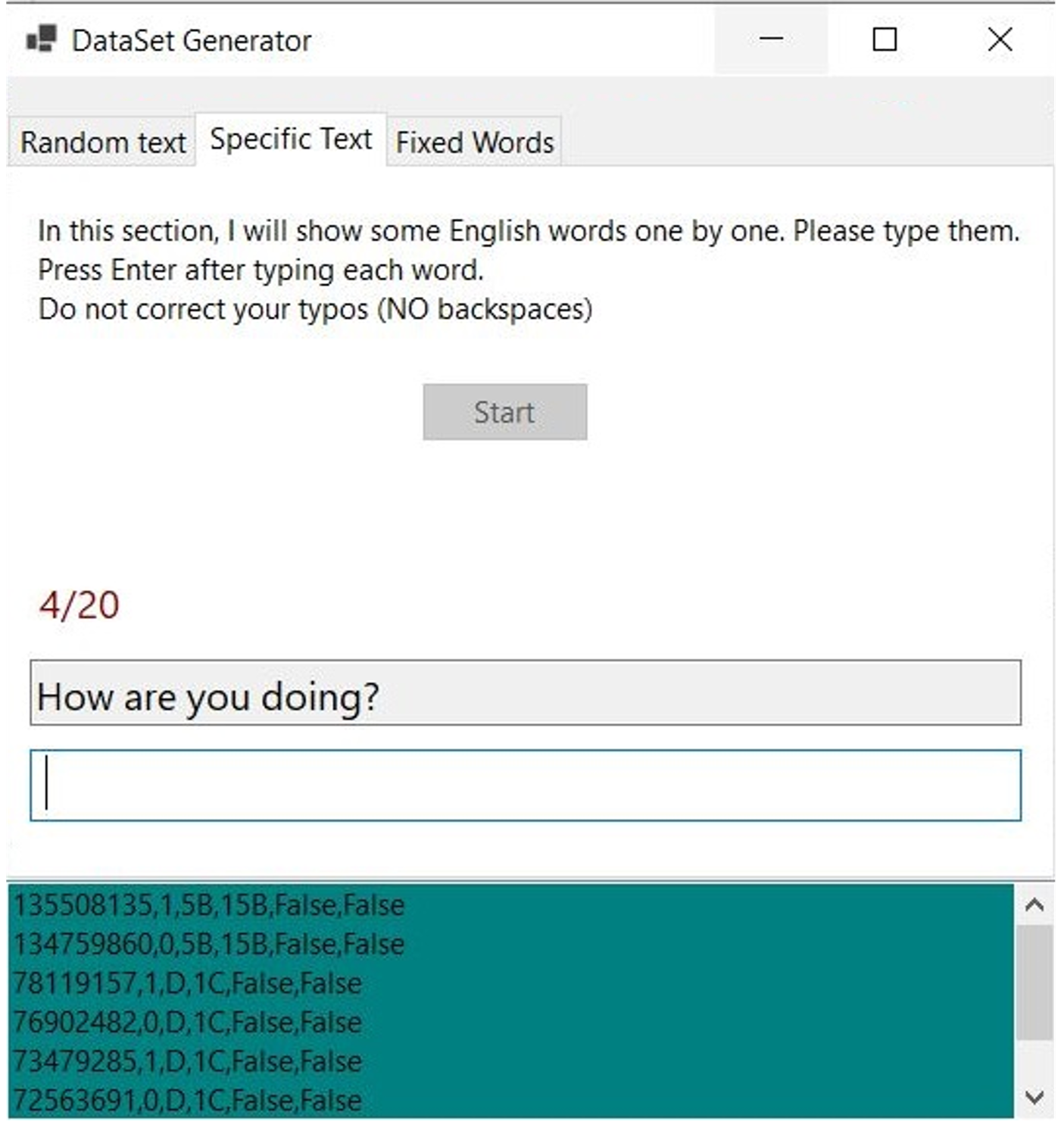}
         \caption{Specific Sentence} \label{fig:text}
     \end{subfigure}
     \hfill
     \begin{subfigure}[b]{0.33\textwidth}
         \centering
         \includegraphics[width=\textwidth]{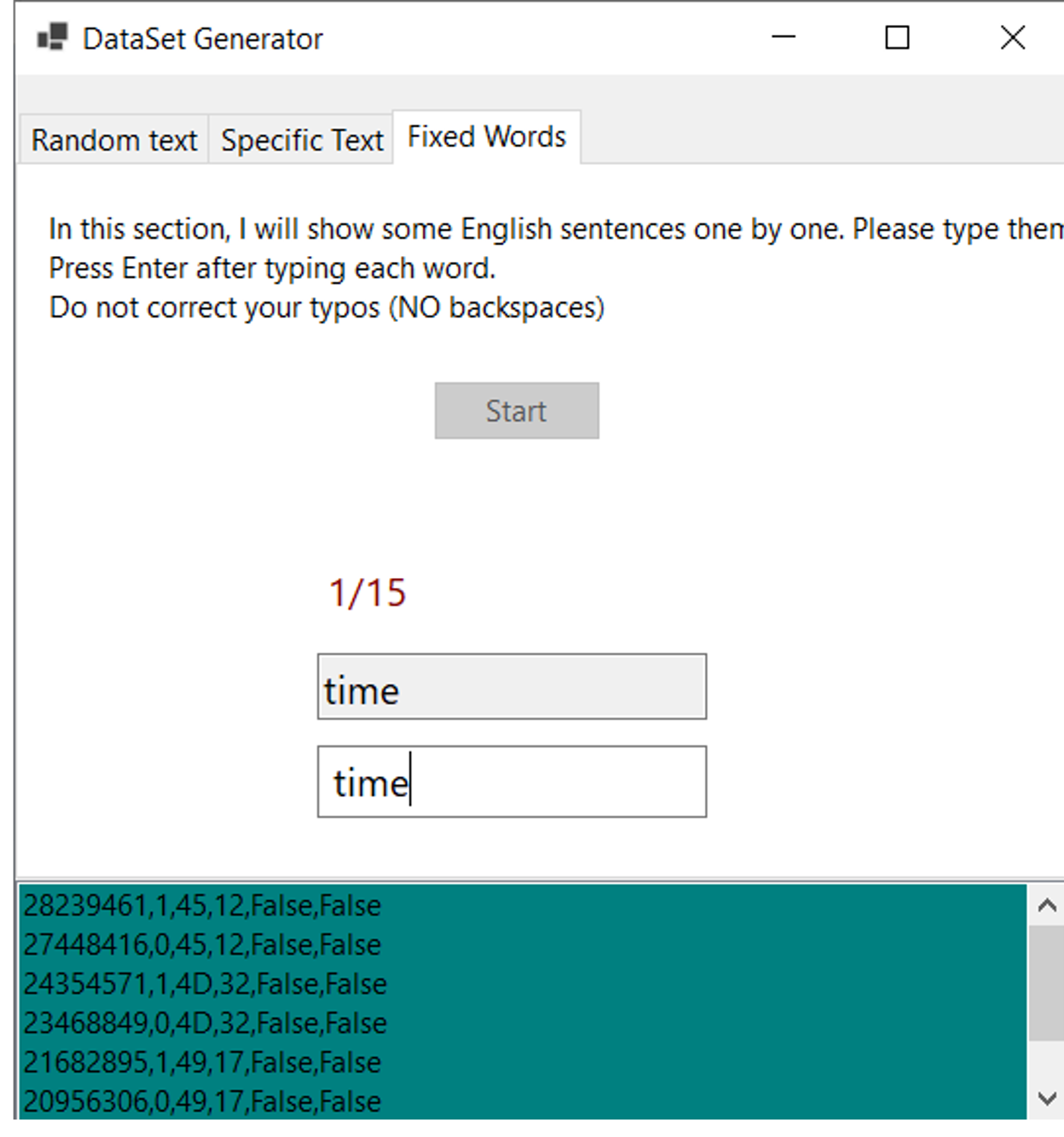}
         \caption{Specific Words} \label{fig:words}
     \end{subfigure}
        \caption{The interface of the data gathering software}
        \label{fig:three graphs}
\end{figure*}

\section{Related Work}

\label{Related Work}
There is a significant body of research on using acoustic side channels to extract information from computing devices. The studies reviewed in this section have made significant contributions to the understanding of keyboard acoustic side-channel attacks. In 2004, Asonov and Agrawal \cite{asonov2004keyboard} demonstrated the possibility of deducing the keystrokes made by an individual on a keyboard by using a simple microphone placed nearby. They achieved this by analyzing the acoustic emissions produced while typing using signal processing techniques. Furthermore, their study revealed the ability to extract sensitive information, such as passwords and credit card numbers, from the sound of keystrokes. 

More recently, in 2011, Genkin et al. \cite{genkin2014rsa} presented a significant attack that used acoustic signals to extract cryptographic keys from a computer's processor. The authors demonstrated the ability to extract RSA keys from laptops, desktops, and even mobile phones using a smartphone placed near the target device. This work showed that even highly secure cryptographic systems can be vulnerable to acoustic side-channel attacks. In addition to attacks on computers, there have been studies on using acoustic side channels to extract information from other devices. For example, Quarta et al. \cite{quarta2018acoustic} showed that it was possible to infer the content of printed documents by analyzing the acoustic signals generated by a printer. They showed that it was possible to reconstruct printed text with up to 90\% accuracy from acoustic signals captured by a smartphone placed near a specific printer.
Another example is the work by O'Leary et al. \cite{o2016acoustic}, who showed that we can use acoustic signals to infer the location of a smartphone in a room. By analyzing the acoustic signals generated by the smartphone's speaker and microphone, the authors were able to estimate the position of the device within a few centimeters. Also, there have been studies on using acoustic side channels to extract information from human-computer interfaces. For example, Backes et al. \cite{backes2013acoustic} showed that it was possible to infer PIN codes entered using a touchscreen by analyzing the sound of the user's finger tapping on the screen. 

In the work by De et al. \cite{de2019differential}, a low-cost side-channel attack called DAA (differential audio analysis) exploited the sound generated by mechanical keypads. This attack, which was based on differential characteristics between sounds recorded by two microphones, achieved a 100\% classification rate for two tested PIN pads of the same model. However, when applied to a different device, the success rate dropped to 63\%. In \cite{toreini2015acoustic}, Toreini et al. (2015) employed signal processing and machine learning techniques to explore the feasibility of discerning individual Enigma keys based on the acoustic emissions they produced. Their results demonstrated that identifying these keys could be consistently achieved with a success rate of 84\%. Notably, this identification process only required basic equipment, including a simple microphone and a personal computer. 
In Anand et al.'s work \cite{anand2018keyboard}, they investigate remote keyboard acoustic side-channel attacks and propose countermeasures. They demonstrate the adaptability of acoustic side-channel attacks to remote eavesdropper settings and introduce a software-based defense, using white noise and fake keystrokes to mitigate these threats effectively.
In Harrison et al.'s recent work \cite{harrison2023practical}, a deep learning model was practically implemented to classify laptop keystrokes using a smartphone microphone. Achieving a remarkable 95\% accuracy with nearby phone-recorded keystrokes and a significant 93\% accuracy with Zoom-recorded keystrokes, this study showcases the feasibility of side-channel attacks using readily available tools and algorithms. The paper also addresses mitigation strategies to safeguard users from these threats.

There are a lot of other approaches that prove the feasibility of using acoustic side channel effects of keyboards to attack sensitive data; however, often, they overlook the effect of the user's typing style in their studies as a critical parameter that can reduce their success rates. In Song et al.'s study \cite{song2001timing}, they conducted a statistical analysis of typing patterns, revealing that such patterns can divulge key information. They utilized a Hidden Markov Model and a key sequence prediction algorithm to predict keystrokes from inter-keystroke timings. Furthermore, they devised an attacker system to monitor SSH sessions for password learning.
In the work by Liu et al. \cite{liu2019human}, the authors introduce a novel approach to user-independent inter-keystroke timing attacks on PINs. Their attack methodology is based on constructing an inter-keystroke timing dictionary derived from a human cognitive model. Significantly, this model's parameters can be determined with a minimal amount of training data from any users, regardless of whether they are the intended target victims. This approach suggests the potential for large-scale deployment in real-world scenarios. The authors explore various online attack settings, evaluating their attack's performance across different levels of PIN strength. Their experimental results reveal that the proposed attack outperforms random guessing methods significantly.


%% file: our_approach.tex
\section{Our Method}

\begin{figure*}
    \includegraphics[width=0.9\textwidth]{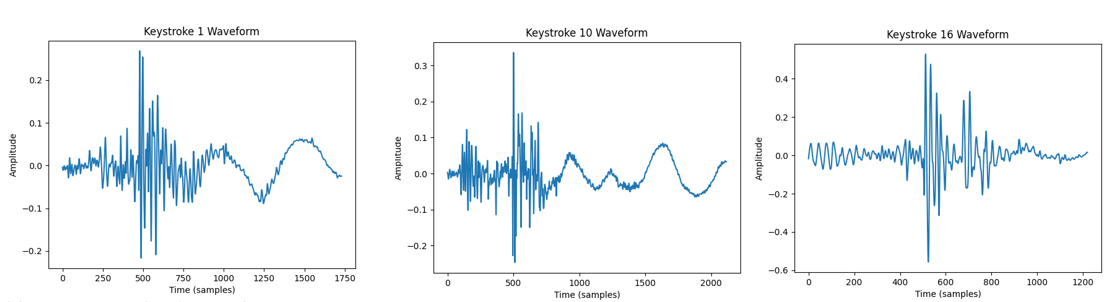}
    \centering
    \caption{Example of extracted segments related to keystrokes}
    \label{fig:segmaents}
\end{figure*}

\label{Implementation}
We propose a new technique that can complement other approaches, covering their weaknesses and defects to achieve better results. Typing pattern is often overlooked by other approaches and negatively affects their results. We turn this negative point into a positive one. The time interval between pressing keys on a keyboard varies for different key pairs. However, we have noticed that an individual tends to type a particular key pair within a similar time interval in most cases, which is influenced by factors such as the arrangement of keys on the keyboard, the anatomy of their hands, and personal typing patterns. For example, we observed that the selected users tended to type the "b" and "o" keys within a similar time in different words, such as "both" and "book," although these time intervals may vary from one user to another. On the other hand, there are variations in the time intervals between different key pairs. Using these two facts, we can model the victim's typing pattern. This model and a recording of the ambient noise of the victim place can lead to a keyboard acoustic side-channel attack. Our method shows that we can deduce sensitive information, such as emails and texts, by analyzing the acoustic emissions produced while typing. In the following sections, we will discuss the different phases of this research in detail.

\subsection{Threat Model}
Our approach assumes the victim is identified. We do not restrict our method to a specific keyboard model or brand. Moreover, we assume that the victim typically works in a private room. Hence, the environment noise is at the level of usual quiet individual offices such that the noises can be controlled and handled through some signal processing methods.
We presume that we can capture some typing samples of the victim in his working environment, including the text and the ambient noise, which we use to build and train our statistical models. 
We consider an oracle that can split the target audio file into separate files corresponding to each word. This is a realistic assumption because usually, after typing a word, users press the Enter or Space keys, which generate distinguishable sounds even by human ears. 

\subsection{Dataset}
To build a dataset of keystroke sounds, we developed a Windows-based application (Figure \ref{fig:three graphs}) using \textit{C\#} programming language\footnote{Python version of this program is available (link to both software available upon request)} that captures a user's typing pattern in three conditions. Firstly, after capturing the ambient noises for the first few seconds, the user types any desired text on their wish. Secondly, he types specific sentences, and in the final step, he types a set of predetermined commonly used words to capture a diverse range of typing styles and patterns. We chose these sentences and words based on the diversity needed to type letters and words in English.

\textbf{Implementation Overview:} When the software is executed, a pop-up appears to inform users about the research policies and to obtain their consent to participate in the study. Subsequently, a new anonymous folder is created within the main Dataset folder. Following this, the software records the keystrokes and audio samples during the three mentioned steps. The resulting dataset consists of three audio files in \textit{WAV} format and three text files for each step. Each text file contains detailed information on each keystroke, including the press and release \textit{TimeSpan}s, \textit{Key status}, \textit{Virtual code}, \textit{Scan code}, \textit{Caps lock status}, and \textit{Shift status}.

The developed software allows us to gather a comprehensive and confidential dataset to capture diverse typing styles and patterns, including the audio noises in the environment. We use the resulting dataset files to train models for our attack.
\subsection{Training Statistical Model}
As mentioned before, the main idea of this paper is to utilize a user's typing pattern to train a statistical model. In the first step, information from the dataset gathered by each user is used to calculate and store the time interval between two pressed keys in a sheet called 'Data' in an Excel file entitled 'model.xlsx'. Next, we calculate the average and standard deviation of these time intervals for each sequence of similar keys and store them in another sheet called 'Analysis' in the same Excel file. The 'Analysis' sheet serves as the trained model.

We utilize the information from both sheets and compare the results. As the number of similar keystroke sequences increases, we can see that the user types those sequences with almost identical time intervals, with only minor differences. Hence, using the 'Analysis' sheet for larger user-generated datasets is preferable. Moreover, by examining the standard deviation values, we can infer the uniformity of the user's typing pattern. The higher the standard deviation values, the more inconsistent the user's typing pattern, resulting in a higher probability of errors in our algorithm. Nevertheless, we propose a coefficient to reduce such errors, which will be discussed in the subsequent sections.

\begin{algorithm}[h]
\caption{Keystrokes Prediction Algorithm}
\label{alg:sound_prediction}
\begin{algorithmic}
\Require  Trained model $M$, Recorded sound of typing $S$, Frame length $l$, Number of keystrokes to predict $k$, Minimum time between two keystrokes $m$, Tolerance factor $t_f$.
\Ensure List of predicted words $W$, list of predicted words that exist in the dictionary $W_c$.
\State Set $F= \{f_i = 0 | 0\leq i<|S|\}$.
\While{$i = 0$ to $|S|-l$}
\State Set $f_i = \sum_{t=i}^{i+l-1} |S[t]|$.
\EndWhile
\While{$n = 1$ to $k$}
\State Set $b_n = Max(F)$.
\State Update $F=\{f_i = 0 |  b_n-m < i < b_n+l+m \}$.
\EndWhile
\State Set $B = (b_1, ... , b_k)$.
\State $Sort(B)$.
\State Set $root = \{'',\{\}\}$ //  $node = \{value,children\}$ 
\State Set $L = \{'a',...,'z'\}$
\While{$i= 1$ to $k-1$}
\State Set $\Delta T_{i} = b_{i+1} - b_{i}$.
\State Set $C_i = \{ (k_a,k_b,\Delta_{ab}) | k_a \in L \land \Delta_{ab} - t_f \leq \Delta T_{i} \leq \Delta_{ab} + t_f \}$
\State Update $L = \{k_b | (k_a,k_b,\Delta_{ab}) \in C_i\} $
\State $AddCandidatesToTree(root, C_i)$
\EndWhile
\State Set $W = GetAllPathesFromRootToLeaves(root)$
\State Set $W_c = [w \in W | w \in Dictionary]$.
\State Return $W_c$.
\end{algorithmic}
\end{algorithm}

\begin{figure*}
    \includegraphics[width=0.55\textwidth]{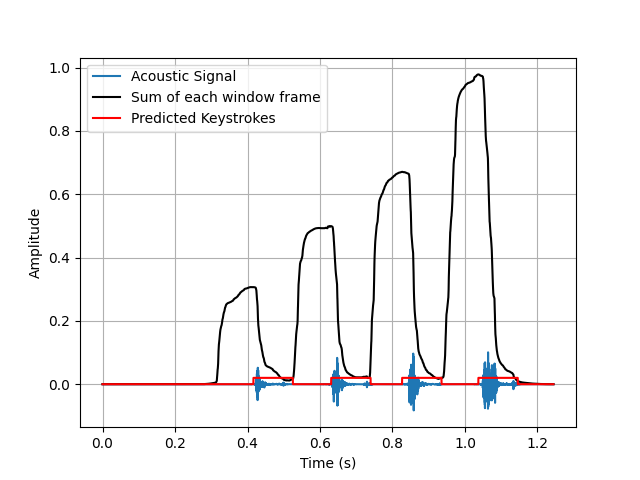}
    \centering
    \caption{Example of detected keystrokes using proposed algorithms}
    \label{fig:detected_keys}
\end{figure*}

\subsection{Prediction}
The next step after determining the statistical model related to each user's typing pattern is to recognize unknown keystrokes by analyzing the audio file recorded from their typing. The audio file needs to be analyzed to find the exact segments related to each keystroke, and then they will be examined through further processes.

\subsubsection{Extracting Keystrokes Segments}
The process of extracting keystroke-related segments (e.g. Figure \ref{fig:segmaents}) from the audio file involves several steps. Previous works have also noted that keystrokes, especially in the hunt-and-peck style, typically occur within a maximum duration of about 100 milliseconds, notably slower than alternative typing methods.
To begin, we employ a sliding window with a frame size of 100 milliseconds and a hop of one, as illustrated in Figure \ref{fig:detected_keys}. Within this window, we calculate the sum of the absolute amplitude of the recorded audio signal.
Subsequently, we systematically search for the maximum point within these sliding windows and reset the values in the vicinity of this maximum to zero. This resetting operation extends over a range that spans from the index of the maximum point minus the frame length to the index of the maximum point plus the frame length. This range alteration allows us to pinpoint the keystroke signals, as we observe that the time intervals between keystrokes tend to be longer than 100 milliseconds. By zeroing this range, we effectively eliminate the possibility of redundant keystroke predictions. We repeat this step iteratively until we've successfully identified the total number of keystrokes. This determination is made by listening to the recorded file and acting as an operator to ensure accuracy in our predictions.

Since the touch and press components of each keystroke generate the loudest sounds, the maximum amplitude often occurs within the initial ten percent of the keystroke segment. As our third step, after identifying all desired maximum indexes associated with each keystroke, we proceed to calculate segments corresponding to the keystrokes within a range extending from these indexes to indexes plus a frame length. These segments can represent the detected keystrokes. Additionally, we can save each segment in separate files, which can facilitate future analysis and enhance the comprehensibility of the extracted information for further study.

\subsubsection{Calculation of time intervals}
As previously mentioned, determining the time intervals between each keystroke can be achieved once we have detected each segment of keystrokes. Various methods were employed to calculate these intervals, including determining the time between the start or end of each segment, finding intervals between the average start and end times of each segment, and identifying the time intervals between the largest peaks within each segment. We opted to calculate the time intervals between the starting points of all keystroke segments. This approach yielded the best results as it maintained the highest consistency with the values recorded in the dataset and the statistical model we developed.

\subsubsection{Finding sequences of possible keystrokes}
In the preceding section, we calculated time intervals between unidentified keystrokes. Now, we seek the closest match within our statistical model for these intervals. However, as users may not consistently type the same key sequence at exact time intervals, we explore numbers within a range around the calculated interval. This range extends approximately five percent greater and five percent less than the calculated interval. We previously mentioned that a higher standard deviation in time values for similar keys can lead to a higher error rate in our algorithm. To address this, we calculate coefficients for the time intervals based on the average standard deviation and adjust the mentioned five percent range accordingly.

Once we've identified all possible sequences of typed letters for each step, we construct a decision tree-like diagram for the subsequent steps. This diagram enables us to trace all branches connected from the root to the final branch. Any remaining unconnected branches are pruned. In the next step, we consolidate all connected branches, representing the potentially typed keystrokes we've identified. These serve as the raw prediction results.

\subsubsection{Enhancing the results using English dictionary}
The previous step generated a set of potential words typed by the user. Given our specific focus on text content, including emails, and notes, we can augment our prediction accuracy by employing semantic correctness approaches, such as dictionary checks or AI-based methods like Large Language Models (LLMs) \cite{brown2020language}. These methods enable us to refine our predictions and ascertain the accurate sentences and text corresponding to our predicted words. So, in the subsequent step, we enhanced our results by searching for these words in an English dictionary and eliminating invalid words. This process significantly improves the accuracy of our term prediction.

The procedure outlined in this section enables us to predict the letters the user has typed by utilizing a model trained on the user's typing pattern and employing a series of analytical procedures on the audio signal. In the following section, we will discuss the results and experiments conducted.

%% file: Experimental_analysis.tex
\section{Experimental Analysis}
\label{Experimental Analysis}
This section provides a comprehensive analysis of the experimental results derived from our investigation. Our experiments were crafted to evaluate the effectiveness and implications of the proposed acoustic side-channel attack. Our evaluation encompasses various facets of the study, including accuracy, efficiency, and security implications.

\begin{figure}
    \includegraphics[width=0.4\textwidth]{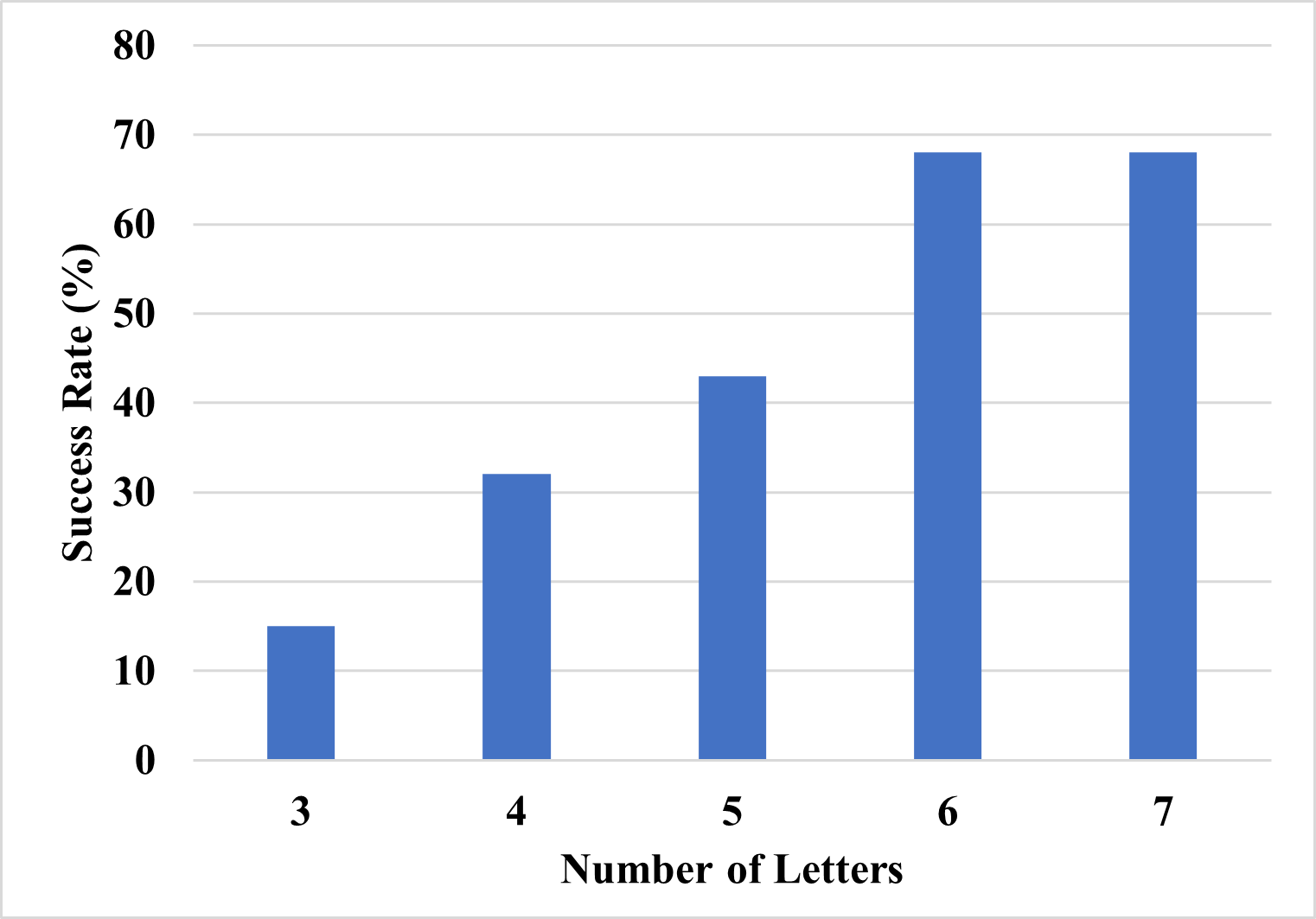}
    \centering
    \caption{The effect of the number of letters on the success rate}
    \label{fig:number_of_letter}
\end{figure}

\subsection{User Study:}
We conducted an IRB-approved user study to gather typing patterns from participants. The data obtained are treated with confidentiality and anonymity. To achieve this, we carefully selected 20 adult users, each exhibiting their unique typing patterns, to train and test the accuracy of our approach using their respective samples. The constructed distinct datasets contained a test set of common English words, such as "work", "love", "life", "like", "night", "world", "table", "they", "have", "teacher", "book", "buy", "credit", "paper", "order", "mobile", "mother", "cat", "run", "house", and "bill", were chosen for their prevalence in English texts and varied word lengths. This selection allowed us to measure the impact of word length on prediction accuracy. The effect of the number of letters in a word is visually represented in Figure \ref{fig:number_of_letter}. The figure indicates that the success rate increases as the number of letters in words increases until we reach six letters. This is because having more letters creates a larger tree based on those letters. However, larger trees do not necessarily result in more meaningful words. Therefore, we have observed that in this situation, there are fewer words that result in bigger success rates. Interestingly, after number six, we reached the highest success rate.

\begin{figure}
    \includegraphics[width=0.4\textwidth]{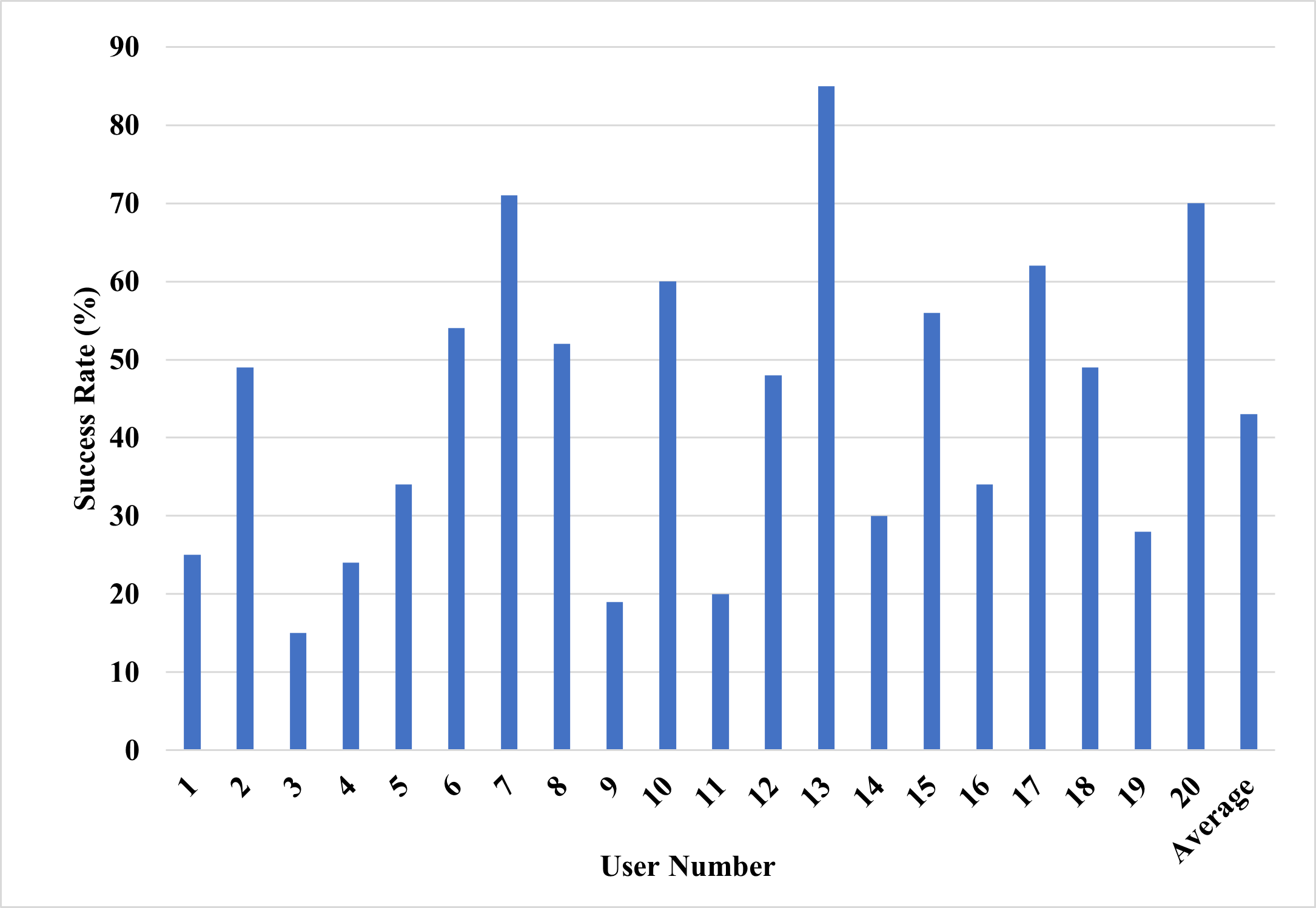}
    \centering
    \caption{The success rate per each user}
    \label{fig:user_accuracy}
\end{figure}

\begin{table*}[h]
\centering
\caption{Comparison of Key Characteristics Among Acoustic Side Channel Attack Approaches on Keyboards}
\begin{tabular}
{|c|p{.6in}|p{.6in}|p{.6in}|p{.7in}|p{.7in}|p{.5in}|}
\hline 
\rowcolor{Gray} \textbf{Approach} & \textbf{Requires large dataset} & \textbf{Specific keyboard dependency} & \textbf{Model training complexity} & \textbf{typing pattern limitation} & \textbf{Environ-ment noise limitation} & \textbf{Success Rate} \\
\hline
cognitive modeling \cite{liu2019human} & No & Yes & Yes & No & Yes & \%10\\
\hline
Deep learning \cite{slater2019robust} & Yes & Yes & Yes & Yes & No & 84.59\% \\
\hline
Hidden Markov Model \cite{foo2010timing} & No & Yes & Yes & Yes & Yes & N/A \\
\hline
triangulation \cite{ranade2009acoustic} & Yes & Yes & Yes & Yes & Yes & 87.5\% \\
\hline
timing dictionary \cite{liu2019keystroke} & No & Yes & Yes & Yes & Yes & 14.0\% \\
\hline
\textbf{Our Method} & \textbf{No} & \textbf{No} & \textbf{No} & \textbf{No} & \textbf{No} & \textbf{43\%}\\
\hline
\end{tabular}
\label{tab:approaches-comparison}
\end{table*}

\subsection{Accuracy Assessment:}
We conducted a series of experiments and calculations to assess the accuracy of our proposed acoustic side-channel attack. The primary objective of these evaluations was to gauge the effectiveness of our approach in predicting user keystrokes based on their typing patterns. To assess the overall accuracy of the proposed method, we initially tallied the number of correctly detected words for each user's test dataset. Following this, we calculated the average number of correct detections for each user. Subsequently, we computed the average accuracy of our approach across all 20 users. Figure \ref{fig:user_accuracy} illustrates both the collective accuracy of all users and the individual accuracy of each user. Our investigation reveals a linear relationship between the average standard deviation (ASD) of each user's typing sample and the success rate of prediction. This implies that users with a consistent typing pattern tend to have lower ASD, resulting in a higher prediction success rate. For instance, User 13 exhibits the lowest ASD and, consequently, the highest success rate.
Our findings substantiate our initial hypothesis regarding the correlation between typing pattern and susceptibility to these types of attacks. Additionally, we computed the average success rate across all users as the final average success rate for our approach, which amounts to approximately 43\%.

As mentioned earlier, our aim was to simulate a highly realistic environment and experiments. We deliberately chose:
\begin{itemize}
  \item Not to eliminate acoustic noises during the data-gathering process
  \item Not to restrict the keyboard model
  \item Not compelled participants to adopt a specific typing style, such as hunt and peck, touch typing, two-fingered typing
  \item To keep our attack simulation more realistic,
  \item To use the laptop's microphone, which was not a high-quality microphone
\end{itemize}
Consequently, comparing our results with other approaches becomes challenging, as they often constrain various parameters in their experiments to showcase the success rate of their methods. However, we tried to compare some of our features and achievements to other similar methods in table \ref{tab:approaches-comparison}.

In comparing various acoustic side-channel attack approaches on keyboards, distinctive characteristics emerge. Cognitive modeling \cite{liu2019human} stands out for not requiring a large dataset and considering both specific keyboard dependency and users' typing patterns; however, it does not account for environmental noise, resulting in a relatively low success rate of 10\%. Deep learning \cite{slater2019robust} demands a large dataset, exhibits keyboard dependency, and involves complex model training. It achieves a success rate of 84.59\%, but it neglects consideration of realistic environments. Hidden Markov Model \cite{foo2010timing} is notable for its lack of dataset requirement; However, we could not find their final achievement accurately, and they overlook both users' typing patterns and environmental noise. Triangulation \cite{ranade2009acoustic} requires a large dataset and is keyboard-dependent, yet it achieves a commendable success rate of 87.5\%, neglecting considerations for users' typing patterns and environmental noise. Timing dictionary \cite{liu2019keystroke} does not necessitate a large dataset but is keyboard-dependent, resulting in a modest success rate of 14\%, with no consideration for users' typing patterns or environmental noise.

Our proposed method differs by not requiring a large dataset and minimizing keyboard dependency. It excels in considering users' typing patterns and environmental noise, achieving a practical success rate of 43\%. This highlights the trade-offs and strengths of each approach, emphasizing the importance of tailored considerations for real-world applicability.
Our intention was to propose a practical approach aligned with real-world scenarios. Unlike approaches that limit destructive parameters for experimental convenience, we sought to mirror the conditions encountered in the actual world. Note that any success rate greater than zero in real-world scenarios is a significant achievement for an attacker. Furthermore, our proposed method can be combined with other approaches to get better results, addressing potential shortcomings they may have.


\section{Limitations and Countermeasures}
\label{Limitations}
We have tried to minimize our approach's dependence on environmental conditions. However, it is a fundamental reality that if we cannot accurately capture the sound of the keyboard, we may not be able to precisely identify every keystroke, resulting in a reduction in overall accuracy. Another physical limitation to consider is that all acoustic keystroke detection methods rely on the strength of the sound generated by the keyboards. Therefore, keyboards with softer keys produce less sound power, which, in turn, can lead to reduced accuracy in the final detection.

A crucial assumption in our approach is that users maintain a consistent and detectable typing pattern throughout the dataset-building process. In other words, our approach can adversely affect the success rate of these groups:

\begin{itemize}
  \item People who use computers rarely, because they do not maintain a consistent typing pattern.
  \item Professional typists, because they type very fast, and the \textit{touch},  \textit{key-press}, and \textit{key-release} events of consecutive keystrokes usually have overlaps, and we can't detect the tying pattern accurately.
\end{itemize}

%% file: Conclusion.tex
\section{Conclusion}
\label{Conclusion}
In this paper, we have explored the possibility of acoustic side-channel attacks on keyboards that are based on the user's typing pattern. This approach addresses some of the limitations of other methods and provides a more realistic method that can be used against criminal and terrorist groups.

Our proposed method uses the time differences between consecutive keystrokes to train a statistical model for extracting unknown keystrokes from a recorded acoustic sound. To test our method, we collected the ambient noise and typed text of 20 people based on an IRB-approved approach, and obtained approximately a 43\% success rate through our experiments. Our results were achieved in a realistic environment without any restrictions on users' keyboards, typing patterns, or environmental acoustic noises, which is a significant advantage over other approaches. We utilized an English dictionary to enhance our text-detection abilities. However, we intend to explore the potential of AI-powered models like LLMs in future projects to improve our success rate.

We acknowledge that our proposed method still has some inherent limitations, such as the consistency and predictability of the user's typing pattern and the quality of the acoustic recorded sound. Overall, we believe that this approach highlights the need for more secure input methods and emphasizes the potential vulnerability of digital security.